\documentclass[traditabstract]{aa}
\usepackage{graphicx}
\usepackage{txfonts}
\usepackage{amssymb}

\begin{document}

\def\mpc{h_{75}^{-1} {\rm{Mpc}}} 
\def\kpc{h_{75}^{-1} {\rm{kpc}}}
\newcommand{\mincir}{\raise
-2.truept\hbox{\rlap{\hbox{$\sim$}}\raise5.truept\hbox{$<$}\ }}
\newcommand{\magcir}{\raise
-2.truept\hbox{\rlap{\hbox{$\sim$}}\raise5.truept\hbox{$>$}\ }}
\newcommand{\ha}{\mathrm{H}\alpha}
\newcommand{\hb}{\mathrm{H}\beta}

\title{The Environment of HII Galaxies revisited}

\author{E. Koulouridis\inst{1}, M. Plionis\inst{2,3},
  R. Ch{\'a}vez\inst{3}, E. Terlevich\inst{3}, R. Terlevich\inst{3,4},
  F. Bresolin\inst{5}, S. Basilakos\inst{6}}

\institute{Institute for Astronomy \& Astrophysics, Space Applications \& Remote Sensing, 
National Observatory of Athens, Palaia Penteli 15236, Athens, Greece.
\and Physics Department of Aristotle University of Thessaloniki,
University Campus, 54124, Thessaloniki, Greece
\and Instituto Nacional de Astrof\'{\i}sica Optica y Electr\'onica, Puebla, C.P. 72840, M\'exico.
\and Institute of Astronomy, University of Cambridge, Madingley Rd, CB3 OHA, Cambridge.
\and Institute for Astronomy of the University of Hawaii, 2680 Woodlawn Drive, 96822 Honolulu, HI, USA
\and Academy of Athens Research Center for Astronomy \& Applied
Mathematics, Soranou Efessiou 4, 11-527 Athens, Greece }

\date{\today}

\abstract{We present a study of the close ($\mincir 200 \kpc$)
  environment of 110 relatively local ($z\mincir 0.16$) HII galaxies,
  selected from the Sloan Digital Sky Survey (SDSS; DR7). We use
  available spectroscopic and photometric redshifts in order to
  investigate the presence of a close and possibly interacting
  companion galaxy. Our aim is to compare the physical  properties of
  isolated and interacting HII galaxies and investigate possible
  systematic effects in their use as cosmological probes. We find that
  interacting HII galaxies tend to be more compact, less luminous and
  have a lower velocity dispersion than isolated ones, in agreement
  with previous studies on smaller samples. However, as we verified,
  these environmental differences do not affect the cosmologically
  important $L_{H\beta}-\sigma$ correlation of the HII galaxies.}

\authorrunning{E. Koulouridis et al.}
\titlerunning{The Environment of HII Galaxies Revisited}
\maketitle

\section{Introduction}
HII galaxies are compact dwarf objects with massive
star formation bursts. They are characterized by a high luminosity per unit
mass, concentrated mostly in a few strong emission lines in the optical
rest frame, a fact that makes them visible at very large redshifts. This,
together with the observed correlation between the
luminosity of recombination lines, e.g. $L(\mathrm{H}\beta)$
and the ionized gas velocity dispersion $\sigma$ (see
Terlevich \& Melnick 1981; Melnick, Terlevich \& Moles 1988;
Fuentes-Masip et al. 2000; Telles et al. 2001; Bosch, Terlevich \& Terlevich 2002;
Siegel et al. 2005; Bordalo and Telles 2011) renders them 
alternative cosmological distance probes. In Plionis et al. (2011) we presented a thorough investigation
of the viability of using HII galaxies to constrain the dark energy
equation of state and they indeed appear to be a prominent
cosmological probe (see also Melnick, Terlevich \& Terlevich 2000; Siegel
et al. 2005). This was clearly verified by using them to estimate
the Hubble constant, finding a value $H_{0}=74.3 \pm 4.3$ km s$^{-1}$
Mpc$^{-1}$ (Ch{\'a}vez et al. 2012), in excellent agreement with, and
independently confirming, the most recent SNIa based results (Riess et
al. 2011; Freedman et al. 2012).

The cosmological importance of the HII galaxies forces us to
investigate all possible sources of systematic effects that could
affect the observed $L(\mathrm{H}\beta) - \sigma$ correlation. One such
systematic could well be related to the effects of the close 
environment of HII galaxies.
It is widely accepted that interactions between two galaxies
are capable of triggering starburst events by driving gas and
molecular clouds from the outskirts toward the center of each galaxy
(e.g. Li et al. 2008; Ellison et al. 2008; Ideue et al. 2012). These
events enhance greatly the star formation rate of the galaxies and
consequently they appear intensely blue because of their abundance in
young stars. The idea of the interactions-starburst connection was
greatly supported by the studies of ultraluminous infrared galaxies
(e.g. Sanders \& Mirabel 1996; Surace et al. 1998) which were found to be strongly interacting
and by definition highly star forming. 
In addition, supportive evidence of interaction-induced star formation,
at lower infrared luminosities, was given by Koulouridis et al. (2006).

On the other hand, regarding HII galaxies, previous studies on the
environmental dependence of their properties  
concluded that they appear to be less clustered than "normal"
galaxies (e.g. Iovino, Melnick \& Shaver 1988; Loveday, Tresse \&
Maddox 1999; Telles \& Maddox 2000) and to have a deficiency in bright
neighbours (Campos-Aguilar \& Moles 1991; Campos-Aguilar, Moles \&
Masegosa 1993). These results questioned the efficiency of
interactions as the starburst's triggering mechanism at least for the
specific objects. However, Noeske et al. (2006) argued that since
$\sim$30\% of their sample's star forming dwarf galaxies (SFDGs) have
mostly dwarf neighbours, this percentage is a lower limit
because of the poor completness of the NED\footnote{NASA/IPAC
  Extragalactic Database is operated by the Jet Propulsion Laboratory, California
Institute of Technology, under contract with the National Aeronautics
and Space Administration.}, 
which they used to conduct their search. In addition, numerical simulations
(Bekki, 2008) also showed that some compact star forming galaxies can
be the result of dwarf-dwarf merging. We should note however, that
although faint dwarf neighbours (in projection) were also probably
found in a sample of SFDGs by Brosch et al. (2006), they were
considered as non-interacting, because of their large distance, and
rather as a sign of synchronized star formation over a large area. 
Interestingly, Telles \& Terlevich (1995) found, by investigating the environment of 51
HII galaxies, that only $\sim$ 10\% of their sample had a luminous neighbouring galaxy
and they tend to be those of lower $\rm H\beta$ luminosity, lower velocity
dispersion and regular morphology, while on the contrary the majority
of luminous objects seem to be irregular, disturbed and isolated (see
also Telles, Melnick \& Terlevich 1997). Similar results were also
reported in Vilchez (1995) where the SFDGs in low density regions have
larger H$\beta$ equivalent widths and higher H$\beta$ luminosities.

Most of these studies however, investigated the effects of what
could be called the large scale environment, since their radial limit for
the identification of a possible neighbour was at least 1 $\mpc$. In 
addition they were relatively "shallow" because of the low magnitude
limit of the available redshift surveys at the time and as a result
they were sensitive only to the more luminous and massive
neighbours.

The aim of the present study is to investigate the
environment of a larger sample of 128 HII galaxies, selected from the
SDSS, which enables us to perform a consistent environmental analysis 
using a fixed magnitude difference between the HII galaxies and their neighbours, while
reaching fainter magnitudes. More importantly, we would like to
investigate the already mentioned trend reported by Telles \&
Terlevich (1995) which, if confirmed, could introduce a systematic
effect in the use of the HII galaxies as cosmological probes.
Throughout our paper we use $H_0=75\; $ km s$^{-1}$ Mpc$^{-1}$.

\section{Sample Selection \& Methodology}
We consider the original sample of 128 HII galaxies, used to
estimate the Hubble constant in Ch{\'a}vez et al. (2012) which was selected
from the SDSS DR7 spectroscopic data within a redshift range
$0.01<z<0.2$. The sources were
 chosen for being compact ($D<5\ \mathrm{arcsec}$) and having 
large Balmer  emission line fluxes and  equivalent widths. 
A lower limit of 50 \AA\ for the $\hb$ equivalent width  
($W$) was chosen  in order to avoid more evolved
starbursts, that would present underlying absorptions due to an older
stellar population component, thus affecting the emission lines flux
[cf. Melnick, Terlevich \& Terlevich (2000)]. 

High resolution echelle spectroscopy was performed at 8 meter class
telescopes (Subaru \& VLT)
and long slit spectrophotometry 
at the 2.1m telescope of the 
Observatorio Astron\'omico Nacional (OAN) in San Pedro
M{\'a}rtir and at the 2.1 m telescope of the
Observatorio Astrof\'isico Guillermo Haro (OAGH) in Cananea, both in Mexico.
Full details of the sample selection, observations and data reduction
and analysis are given elsewhere (Ch\'avez et al., 2012; Ch\'avez et
al. 2013 in preparation). 

In order to identify neighbours around each HII galaxy, within the
Sloan Digital Sky Survey (SDSS; DR7), we apply a projected rest-frame
maximum radius separation of $<200 \kpc$, as well as radial velocity
limit separation of $\Delta u<600$ km/sec (similar to the
pairwise galaxy velocity dispersion; e.g. Jing, Mo \& Boerner 1998), when spectroscopic
redshifts are available, or $\Delta z<0.025$ (ie., the rms error of
the SDSS photometric $z$'s) when only photometry is available. 
Even though there is no general concensus on the maximum radial
separation of a galaxy pair, most of the recent studies use a search
radius between $20 h^{-1}$kpc (e.g. Patton et al. 2005) and $200
h^{-1}$kpc (e.g. Focardi et al. 2006; see also relevant discussion in
Deng et al. 2008). We choose the limit of 200 $\kpc$ considering that
it is a reasonable distance for a satellite galaxy in a massive halo
(e.g. Bahcall et al. 1995; Zaritsky et al. 1997), while this limit is
also a compromise between having enough ``isolated'' and ``paired''
HII galaxies. Had we increased its value we would reduce greatly the
number of isolated galaxies and vice versa.

In addition, we limit our neighbour search to a maximum SDSS $m_{r}$-band magnitude
difference between the HII galaxy and its companion of $\Delta m_{r}=+1$
since the SDSS completness limit is $m_{r}\sim 20.5$  while
our HII galaxy sample is limited to $m_{r}<19$, reducing our sample to 110 objects.

Our aim is to separate the isolated HII galaxies from those with at least
one neighbour within the specified angular and velocity limits, and
then compare their physical
properties, i.e. their velocity dispersion $\sigma$, $\rm H\beta$ emission
line luminosity\footnote{Throughout when referring to
  luminosity we will always imply $\rm H\beta$ luminosity.} $L_{H\beta}$ and
metallicity $Z$ (defined as O/H abundance). 

In our initial analysis we choose to use only the HII galaxies which have
neighbours with spectroscopic redshifts (ie., we
exclude all HII galaxies that have neighbours based only on
photometric $z$'s, due to the known relatively large photo-$z$
uncertainty).
Note that although the above exclusion of photo-$z$ pairs is expected to
reduce the noise in our results, it could introduce a bias
towards brighter pairs since the spectroscopic SDSS catalogue is
complete only to an $r$ magnitude of $17.77$. This bias manifests itself also as a redshift
distribution difference between the HII galaxies having a neighbour
and the isolated ones. Nevertheless, we will use these results as a
starting point. Ideally, we would like to compare
subsamples matched in redshift and with available spectroscopy. This
is possible at present only for a redshift limited sample of $z=0.05$,
below which there is no redshift distribution difference of the two
subsamples. However, in this case the size of the sample is greatly
reduced which may affect the significance of our results.

Therefore, we choose also to use the photo-$z$ based pairs in our
analysis,
since their inclusion (at the expense of additional noise) eliminates
the problem of the uneven redshift distribution between isolated and
paired HII galaxies without reducing the respective subsamples. Our
goal for the future is to obtain spectroscopic redshifts of all the
photo-$z$ based neighbours and confirm our results.

\begin{figure}
\centering
\resizebox{10cm}{10cm}{\includegraphics{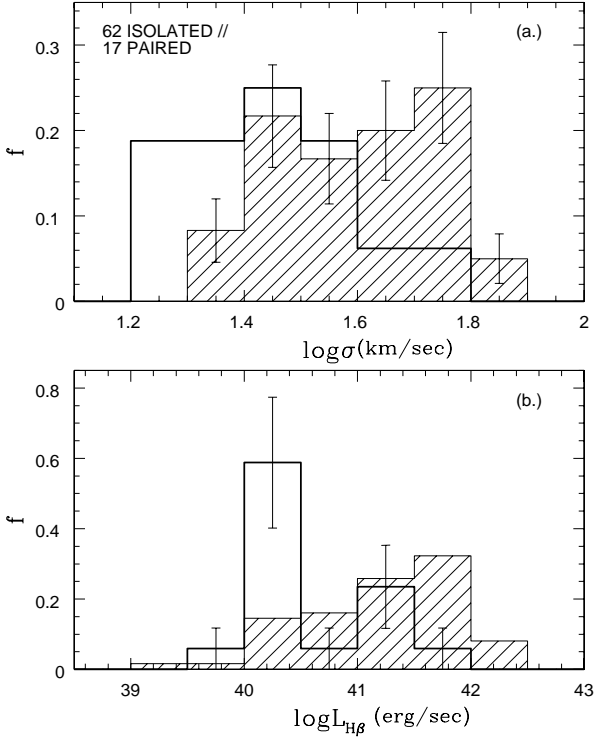}}
\caption{Velocity dispersion (panel a.) and luminosity (panel b.)
  distributions of HII galaxies with (plain) and without (hatched)
  spectroscopically confirmed neighbours. Uncertainties are 1$\sigma$ Poisson 
  errors.}
\end{figure}

Telles \& Televich (1995) concluded that their sample of HII galaxies
with a close companion tends to be more compact and less disturbed
than the respective sample of isolated ones. Because of the faint
magnitudes and compactness of the majority of our sample galaxies we
are not able to reach a definite conclusion on the latter.
We will however investigate the possible role of interactions in the
compactness of the HII galaxies.
To this end we use the physical diameters of the HII galaxies which are derived from the
apparent SDSS isophotal diameters (at 25.0 $r$-mag arcsec$^{-2}$). The
data and calculation method are available in the
NED. We should note here
that the Petrosian radius is a better measure of the size of extended
sources, but for compact objects %the comparison of our subsamples 
the isophotal diameters can also be used. To test the different
definitions, we compared the Petrosian to the isophotal diameters of a
small random subsample of our HII galaxies and
found that they are completely consistent.

\section{Results}
\begin{figure}
\resizebox{10cm}{10cm}{\includegraphics{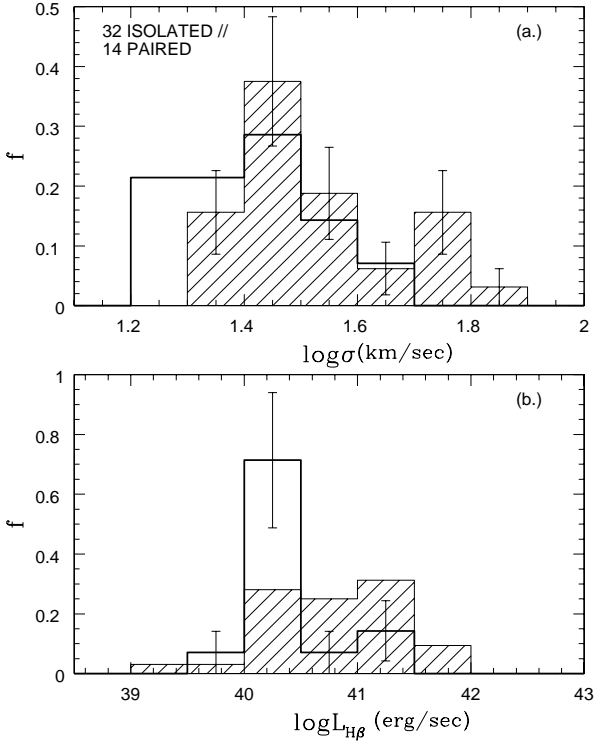}}
\caption{Velocity dispersion (panel a.) and luminosity (panel b.)
  distributions of the volume limited HII galaxy sample ($z<0.05$) with (plain) and without
  spectroscopically confirmed neighbours. Uncertainties are 1
  $\sigma$ Poisson errors.}
\end{figure}

In Fig.1 we plot the luminosity and
velocity dispersion distributions of 62 isolated and 17 paired (using
spec-$z$'s and $\Delta m_{r}=+1.0$) 
HII galaxies. The KS test indicates that the luminosity
and velocity dispersion distributions are significantly different, the
former more than the latter (see Table 1), in the sense that isolated
HII galaxies exhibit higher luminosities and higher velocity
dispersions with respect to those having close neighbours. 
Considering neighbours with $z<0.05$ (Fig.2) supresses the problem
of the uneven redshift distribution between isolated and paired HII
galaxies (last column of Table 1) at the cost of reducing greatly the
number of isolated HII galaxies. The results remain practically the
same, even though less statistically significant (especially for the
velocity dispersion). We should note however that this is probably due
to the small number of available objects.

By considering in the previous analysis only the spec-$z$ based pairs,
in order to avoid projection effects due to unreliable photometric redshifts,
we have excluded more than half
our HII galaxy sample resulting in less significant statistical results.
Furthermore, by not imposing a redshift limit in Fig.1, we may
have also introduced a bias towards brighter HII galaxy neighbours, as we have discussed earlier.
\begin{figure}
\resizebox{10cm}{10cm}{\includegraphics{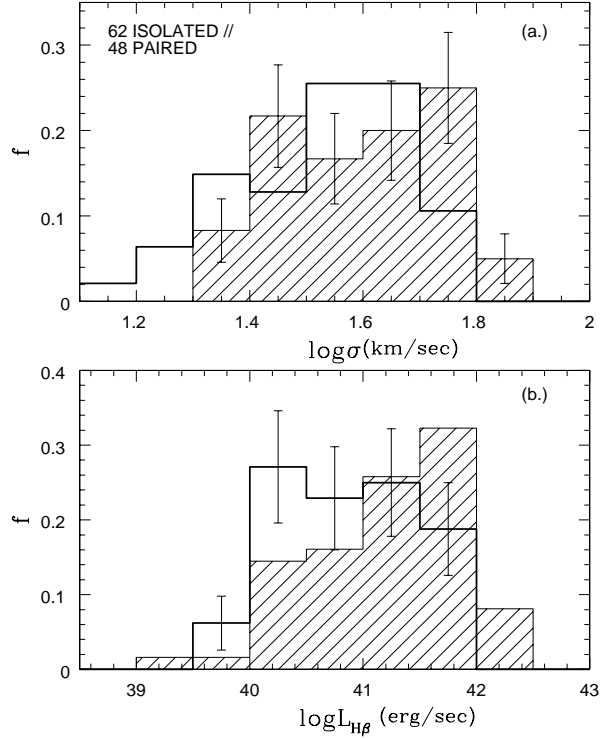}}
\caption{Velocity dispersion (panel a.) and luminosity (panel b.)
  distributions of HII galaxies with (plain) and without (hatched)
  spectroscopically or photometrically confirmed
  neighbours. Uncertainties are 1 $\sigma$ Poisson errors.}
\end{figure}

We now relax the above conditions and we add neighbours with available photometric redshift
within the limit $\Delta m_{r}=+1.0$, considering all
photometrically confirmed neighbours as "true" neighbours (Fig.3).
This increases considerably (almost triples) the
number of paired HII galaxies (from 17 to 48). Despite
the fact that we surely contaminate the subsample of paired HII galaxies 
with a number of isolated ones because of the greater
photometric redshift uncertainty of faint objects, we can
see that the luminosity and velocity dispersion distributions can be
considered statistically different, but the former at a reduced 
confidence level with respect to the Fig.1's case (see Table 1).
In addition, we cannot reject the hypothesis that the redshift
distributions of the two subsamples are drawn from the same parent
population at any significant statistical level. 

We should note here that excluding all galaxies with redshift above
$z=0.05$ from the analysis, presuming that the photometric redshifts
of low $z$ galaxies are more reliable, the trends remain practically
the same as for the whole sample.

In all cases the metallicity distributions of the two subsamples are
statistically equivalent and therefore we do not present any extra
plots.

\begin{table}
\centering
\caption{Results of Kolmogorov-Smirnov statistical tests.}
\tabcolsep 5pt
\begin{tabular}{cccccccc}
\hline
{\em z confirmation} &{\em Fig.}& z &{\em $N_{i}$} & {\em $N_{p}$}&{\em KS$_\sigma$}&{\em KS$_L$}&{\em KSz}\\
{\em (1)} &{\em (2)} & {\em (3)}&{\em (4)}&{\em (5)}&{\em (6)}&{\em (7)}&{\em (8)}\\
\hline
spectroscopy &1&0.15&62&17&0.011&0.003&0.005\\
spectroscopy &2&0.05&32&14&0.098&0.030&0.129\\
spec. \& photometry&3&0.15&62&48&0.067&0.011&0.952\\
\hline
\end{tabular} 
\tablefoot{{\it(1)} Redshift confirmation, {\it(2)} respective figure
  number, {\it(3)} upper limit of HII galaxy redshift, {\it(4)} Number
  of isolated HII galaxies, {\it(5)} Number of HII galaxies with at
  least one neighbour within 200 $\kpc$, {\it(6)} probability that the null
  hypothesis that the samples are drawn from the same parent
  population can not be rejected for velocity dispersion
  distributions, {\it(7)} same as column {\it 6} for luminosity
  distributions, {\it(8)} same as column {\it 6} for redshift
  distributions.
}
 \end{table}  

Returning to the issue of the compactness of the HII galaxies we apply a KS test
on the diameter distributions of a low redshift volume limited subsample, where
all pairs are spectroscopically confirmed (Fig.4a), and of the whole parent sample
using in addition photometrically confirmed neighbours (Fig.4b). Once more, we use
volume limited subsamples to avoid introducing any distance dependent bias. We
conclude that although we cannot reject the null hypothesis that the two
aforementioned distributions are drawn from the same parent population at any
significant confidence level, we observe a difference between the two subsample
distributions which mainly arises from a visible shift between the distributions.
This shift is due to the more compact objects being mostly HII galaxies with
neighbours, while the more extended ones are isolated. We should note here that by
investigating the diameter distributions of the HII galaxies in a small number of
narrow redshift bins, we find that this effect persists in each bin.  This
partially confirms the results of Telles \& Terlevich (1995)  who found a weak
trend where the most compact HII galaxies tend to have a close neighbour, whereas
the most extended ones tend to be isolated. However, this trend between the two
subsamples is not highly significant and a larger sample would be necessary in order 
to confirm this trend.

\begin{figure}
\centering
\resizebox{7cm}{6cm}{\includegraphics{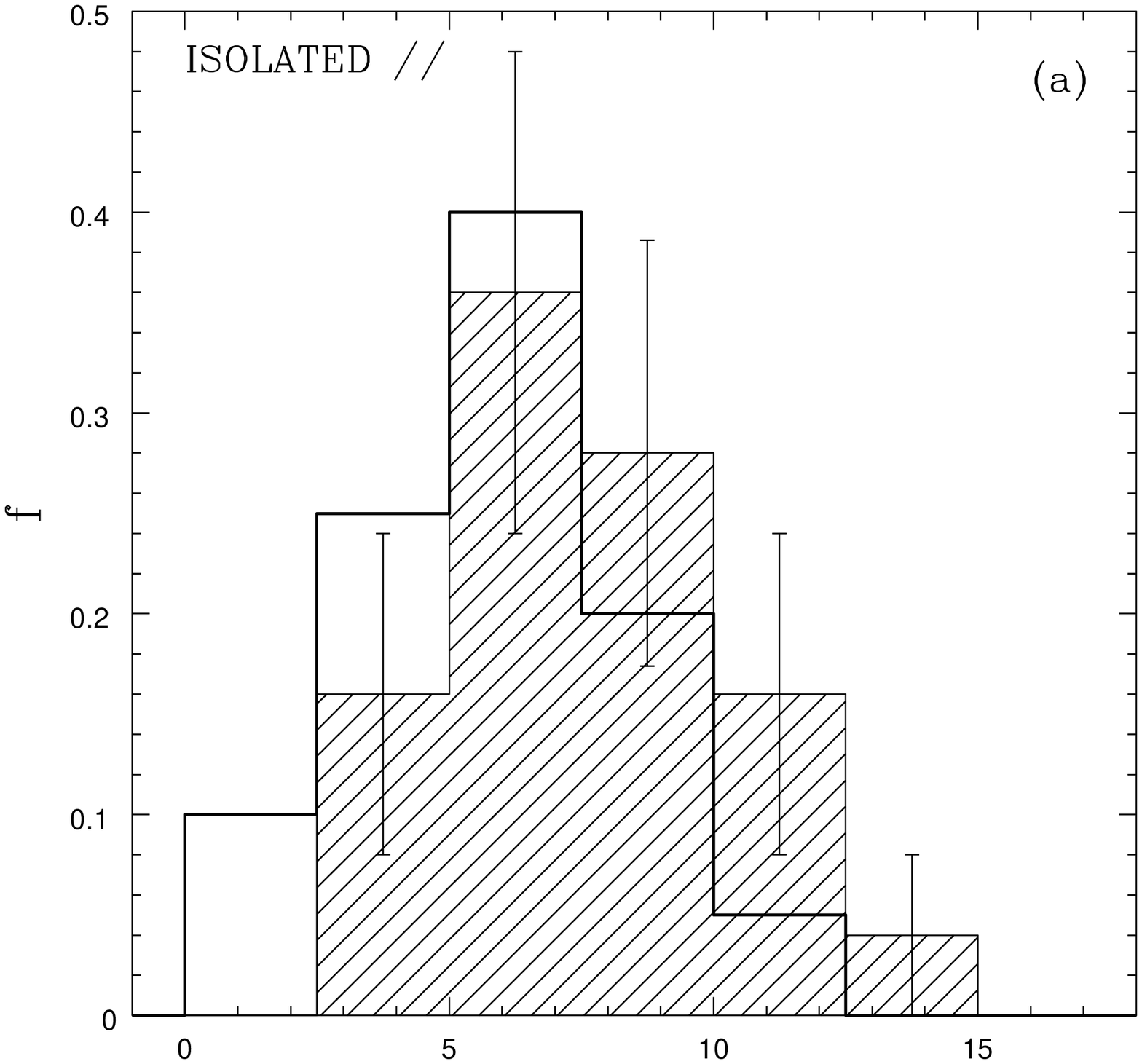}}
\resizebox{7cm}{6cm}{\includegraphics{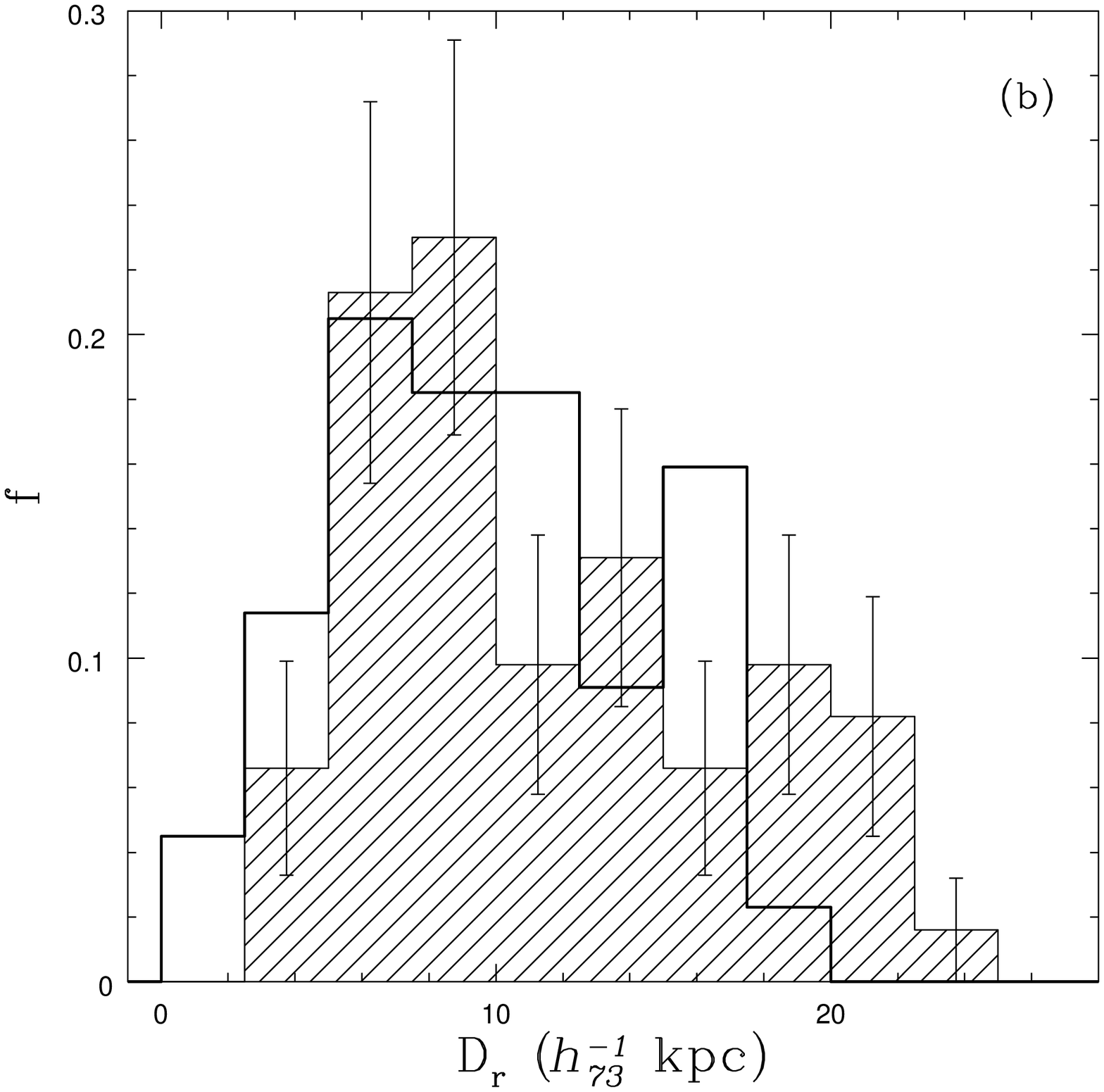}}
\caption{Isophotal r-SDSS diameter distribution of isolated (hatched
  area) and paired HII galaxies of (a) a low redshift ($z<0.05$)
  volume limited subsample of HII galaxies with spectroscopically
  confirmed neighbours, and (b) of the whole parent sample using also
  photometrically confirmed neighbours. Uncertainties are 1$\sigma$
  Poisson errors.}
\end{figure}
We now wish to investigate whether the particular systematic effect
that we have identified, ie., the environmental dependence of both the $H\beta$
luminosity and the velocity dispersion, affects the cosmologically
relevant $L_{H\beta}-\sigma$ correlation. Qualitatively one should not
expect any important effect on the correlation since the existence of
a close neighbour affects both the $H\beta$ luminosity and $\sigma$ in
the same direction; ie., they both decrease following the monotonic
trend of the observed correlation. Nevertheless, in order to
be more quantitative we fit the above correlation separately for the
isolated or the paired HII galaxies, and we compare the slopes of the
correlation with that of the whole sample together. In Fig.5 we plot
the three lines which best fit each sample (black line for the parent
sample (all points), red line for the paired HII sample (open squares)
and green line (triangles) for the isolated sample). The hatched
lines define the confidence band at 2$\sigma$ confidence level of the parent sample's regression
line, taking into account the joint distribution of the slope and the intercept.

For the case of isolated HII galaxies we
find a relative difference of the slopes of the  $L_{H\beta}-\sigma$
relation $\delta a/a_{T}=(a_{isol}-a_{T})/a_{T}=-0.06\pm 0.10$, while
for the case of paired HII galaxies, $\delta a/a_{T}=0.08\pm 0.13$,
where $a_{T}$ is the slope of the correlation based on the parent
sample of HII galaxies. 

In an attempt to determine if any discrepancies can also arise due to
random sampling of 62 or 48 objects drawn from the parent population
independently of their properties, the errors were calculated by
applying the bootstrap method i.e. by resampling randomly each
subsample from the parent sample.
Indeed, in both cases, we find no significant difference. 

This is 
indeed a very important result, indicating that although environmental effects
do influence the dynamics of the starburst they do not affect the
cosmologically important $L_{H\beta}-\sigma$ correlation. Given the size of the sample, the possibility remains that
these small differences between the slopes of the two subsamples are
intrinsic. However, to verify if such small slope
differences are real, we would need at least to triple the
number of objects, a difficult task considering the expensive
observational requirements for this kind of studies. Furthermore, 
such a difference in the slope of the relation is not found even at much
higher densities, as shown by the similarity of the slope of the L-sigma relation for
Giant HII regions in the disks of massive spirals (see conclusions \& discussion section).

\begin{figure}
\resizebox{8cm}{8cm}{\includegraphics{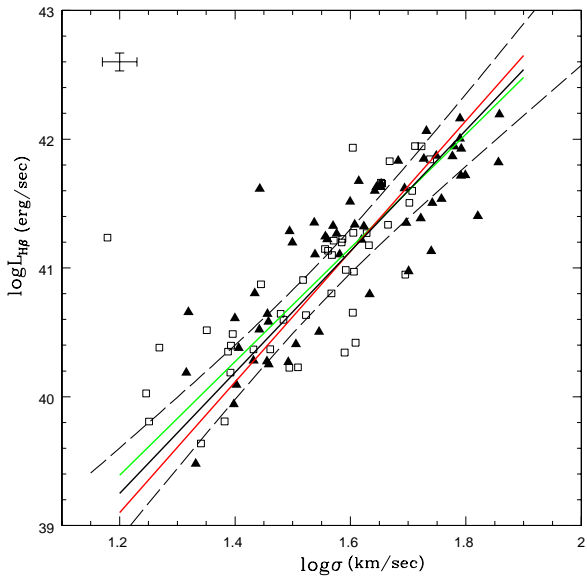}}
\caption{L-$\sigma_{H\beta}$ diagram. The red line denotes the linear regression fitting
  of the isolated HII galaxies (triangles), while the green line
  denotes the fitting of those that have at least one
  spectroscopically or photometrically confirmed neighbour within 200
  $\kpc$ (squares). The black line is the fitting of the parent sample
  and the hatched lines define the 2$\sigma$ confidence band of the regression line. Typical 1$\sigma$ 
  errors are shown on the upper left corner of the plot.}
\end{figure}

\section{Conclusions \& discussion}

We have studied a sample of 110 HII galaxies which was selected
from the SDSS DR7 spectroscopic data within a redshift range
$0.01<z<0.16$ and $m_r<19$. Our 
results indicate that there is a connection between the existence of a companion
and the size of the starforming region.

In particular we find that both The $\rm H\beta$ luminosities and velocity dispersions of the
HII galaxies with neighbours (within a projected rest-frame radius
separation of $<200 \kpc$ and radial velocity separation of $\Delta
u< 600 km/sec$) tend to be significantly lower than those of the
isolated ones.
 
Importantly, the $L_{H\beta}-\sigma$ correlation and distance estimator  is not affected by
the environmental or host galaxy differences of the isolated and paired HII
galaxies. One would like to understand the physical mechanism by which
the environmental dependence of $L_{H\beta}$ and $\sigma$ is such as
to leave unaffected the correlation. If this correlation is a
manifestation of the virial theorem, it is evident that the mass of
the molecular cloud, progenitor of the massive
starburst, is influenced by the environment, i.e. more massive clouds, producing
more massive starbursts and therefore higher values of $L_{H\beta}$ and $\sigma$, are developed in
lower density environments, while the lower mass clouds are characteristic of
higher density environments, where more frequent interactions can limit the growth
of the molecular clouds.
This is even more evident when including in the discussion
the results  of Giant HII regions in the disks of massive spirals. Giant HII
regions are massive bursts of star formation in a much higher density environment
than that of HII galaxies. As can be seen from the analysis of Melnick, Terlevich
\& Moles (1988) or  Ch{\'a}vez et al. (2012), Giant HII regions and HII galaxies
define a tight $L_{H\beta}-\sigma$ correlation where Giant HII regions occupy the lower end of
the relation; i.e. they tend to be much smaller than HII galaxies while
independently verifying the same tight relation between $L_{H\beta}$ and $\sigma$ of the more 
luminous HII galaxies notwithstanding the fact that they are formed in a much
denser environment. 

Our results are in agreement with previous studies showing
that HII galaxies are less clustered than normal galaxies
(e.g. Iovino, Melnick \& Shaver 1988; Loveday, Tresse \& Maddox 1999;
Telles \& Maddox 2000), and that they lack massive neighbours
(Campos-Aguilar \& Moles 1991; Campos-Aguilar, Moles \& Masegosa
1993). Even considering all photometrically confirmed neighbours
(faint neighbours) as real neighbours, more than half of our sample
galaxies remain isolated. Our results therefore concur
with the view where star formation in HII galaxies is not necessarilly
triggered by interactions (Telles \& Terlevich 1995; Telles \& Maddox 2000);
they do appear however to play an important role in the confinement of
the total mass of the progenitor molecular cloud that gives birth to
an HII galaxy. Although the triggering mechanism of the enhanced
star-forming activity of HII galaxies is still debated, star formation 
is probably bound to happen when even
an isolated molecular cloud fulfills the requirements (see discussion
in Telles 2010). Alternatively, and since older stellar populations
have been found
to be present (e.g. Papaderos et al. 1996; Telles \& Terlevich 1997;
Cair\'os et al. 2003), implying that not all HII galaxies are young
formations, it could also be a manifestation of cosmic downsizing;
less massive structures are unable to efficiently form stars in the
past and they are doing so in later epochs, when most massive
structures are already quiescent (e.g. Neistein, van den Bosch \&
Dekel 2006). Thus, given that star formation in most HII galaxies
happens spontaneously and depends only on the mass of the already
virialised system, the $L_{H\beta}-\sigma$ relation should be expected
as well.

Not all HII galaxies should be expected to
verify the $L_{H\beta}-\sigma$ relation with a small scatter. Already
Melnick et al. (1988) had recognised that systems with $\log\sigma >
1.75$ show a flattening of the relation, probably indicating the onset
of rotation for larger starforming regions, and that limiting the
sample to objects with 10 km/s $< \sigma <$ 60 km/s, equivalent widths
$W_{H\beta} > 50 \AA$ and gaussian profiles in their emission lines,
produces a tight $L_{H\beta}-\sigma$ relation, suggesting that we are
dealing with young massive bursts that dominate the luminosity of the
galaxy and that they are gravitationally bound and pressure supported. 
The biases introduced by multiplicity, rotation and contamination
by the underlying galaxy (e.g. Overzier et al. 2008; Amorin et
al. 2012) are
minimised by selecting only objects with emission lines of high
equivalent width and line profiles that are well fitted by a single
gaussian.

\end{document}